\documentclass[prd,amsmath,amssymb,showpacs,superscriptaddress,nofootinbib,twocolumn]{revtex4-1}
\usepackage{multirow}
\usepackage{float}
\usepackage{graphicx}
\usepackage{dcolumn}
\usepackage{bm}
\usepackage{rotating}
\usepackage{color}
\usepackage{subfigure}
\usepackage{pstricks}
\usepackage{soul}
\usepackage{overpic}
\usepackage[english]{babel}
\usepackage[colorlinks,linkcolor=blue,anchorcolor=blue,citecolor=blue]{hyperref}
\usepackage{tabularx}

\newcommand{\jpsi}{J/\psi}
\newcommand{\pip}{\pi^+}
\newcommand{\pim}{\pi^-}
\newcommand{\piz}{\pi^0}
\newcommand{\pp}{\pi^+\pi^-}

\newcommand{\ppbar}{p\bar{p}}
\newcommand{\kskl}{K^0_S K^0_L}
\newcommand{\ks}{K^0_S}
\newcommand{\kl}{K^0_L}

\newcommand{\kk}{K^+K^-}
\newcommand{\kkb}{K\bar{K}}
\newcommand{\Lam}{\Lambda}
\newcommand{\Lamb}{\bar{\Lambda}}

\def \stat{\mbox{$\,$(stat)}}
\def \syst{\mbox{$\,$(syst)}}
\def \model{\mbox{$\,$(model)}}

\begin{document}


\title{\boldmath Study of $\jpsi\to\ppbar\phi$ at BESIII}

\author{
  \begin{small}
    \begin{center}
      M.~Ablikim$^{1}$, M.~N.~Achasov$^{9,e}$, X.~C.~Ai$^{1}$,
      O.~Albayrak$^{5}$, M.~Albrecht$^{4}$, D.~J.~Ambrose$^{44}$,
      A.~Amoroso$^{49A,49C}$, F.~F.~An$^{1}$, Q.~An$^{46,a}$,
      J.~Z.~Bai$^{1}$
% ...
      \\
      \vspace{0.2cm}
      (BESIII Collaboration)\\
      \vspace{0.2cm} {\it
        $^{1}$ Institute of High Energy Physics, Beijing 100049, People's Republic of China\\
% ...
$^{4}$ Bochum Ruhr-University, D-44780 Bochum, Germany\\
        $^{5}$ Carnegie Mellon University, Pittsburgh, Pennsylvania 15213, USA\\
% ...
$^{9}$ G.I. Budker Institute of Nuclear Physics SB RAS (BINP), Novosibirsk 630090, Russia\\
% ...
$^{44}$ University of Rochester, Rochester, New York 14627, USA\\
% ...
$^{46}$ University of Science and Technology of China, Hefei 230026, People's Republic of China\\
% ...
$^{49}$ (A)University of Turin, I-10125, Turin, Italy; (B)University of Eastern Piedmont, I-15121, Alessandria, Italy; (C)INFN, I-10125, Turin, Italy\\
% ...
$^{a}$ Also at State Key Laboratory of Particle Detection and Electronics, Beijing 100049, Hefei 230026, People's Republic of China\\
% ...
$^{e}$ Also at the Novosibirsk State University, Novosibirsk, 630090, Russia\\
% ...
      } \end{center}
  \end{small}
}

\date{\today}

\begin{abstract}
 Using a data sample of $1.31 \times 10^{9}$ $\jpsi$ events
 accumulated with the BESIII detector, the decay $\jpsi\to\ppbar\phi$
 is studied via two decay modes, $\phi\to\kskl$ and $\phi\to\kk$. The
 branching fraction of $\jpsi\to\ppbar\phi$ is measured to be
 $\mathcal{B}(\jpsi\to\ppbar\phi)=[5.23\pm0.06\stat\pm0.33\syst]\times10^{-5}$,
 which agrees well with a previously published measurement, but with a
 significantly improved precision. No evident enhancement near the
 $\ppbar$ mass threshold, denoted as $X(\ppbar)$, is observed, and the
 upper limit on the branching fraction of $\jpsi\to
 X(\ppbar)\phi\to\ppbar\phi$ is determined to be $\mathcal{B}(\jpsi\to
 X(\ppbar)\phi\to\ppbar\phi)<2.1\times10^{-7}$ at the 90\% confidence
 level.
\end{abstract}

\maketitle

\section{\boldmath Introduction}

In 2003, a strong enhancement near the $\ppbar$ mass threshold, known
as the $X(\ppbar)$, was first observed by the BESII experiment in the
radiative decay $\jpsi\to\gamma\ppbar$~\cite{BESII1}. It was later
confirmed by the CLEO and BESIII
experiments~\cite{CLEO1,BESIII1,BESIII2}. Strikingly, no
corresponding enhancements were observed either in
$\Upsilon(1S)\to\gamma\ppbar$~\cite{gamppbarcleo} radiative decays or
in hadronic decays of vector charmonium states below the open-charm
threshold, $e.g.$
$\jpsi(\psi(3686))\to\piz\ppbar$~\cite{BESII1,psippi0ppbarbes} and
$\jpsi\to\omega\ppbar$~\cite{ppomegabes2,ppomegabes3}.

The experimental observations of the $X(\ppbar)$ structure in
$\jpsi\to\gamma\ppbar$ and the absence in other probes raised many
discussions in the community resulting in various speculations on its
nature. The most popular theoretical interpretations include
baryonium~\cite{baryon1,baryon2,baryon3}, a multiquark
state~\cite{muquark} or an effect mainly due to pure final-state interaction
(FSI)~\cite{ppfsi1,ppfsi2,ppfsi3,ppfsi4}. In accordance with the latest
results of a partial wave analysis (PWA)~\cite{BESIII2}, it was
proposed to associate this enhancement with a new resonance,
$X(1835)$, that was observed in the
$\jpsi\to\gamma\pip\pim\eta^{\prime}$ decay~\cite{x18351,x18352}. The
nature of the $X(\ppbar)$ is still mysterious to date, therefore its
investigation via other $\jpsi$ decay modes may shed light on its
nature. The decay $\jpsi\to\ppbar\phi$ restricts the isospin of the
$\ppbar$ system and is helpful to clarify the role of the $\ppbar$ FSI.

In this paper, we report on a search for a near-threshold enhancement
in the $p\bar{p}$ mass spectrum and the possible $p\phi$ ($\bar{p}\phi$)
resonances in the process $\jpsi\to\ppbar\phi$. The decay
$\jpsi\to\ppbar\phi$ was investigated by the DM2 Collaboration
based on $(8.6\pm1.3)\times10^{6}$ $\jpsi$ events about thirty years
ago~\cite{DM2}, with a large uncertainty due to the limited
statistics (only $17\pm5$ events were observed). In this work, the
channel $\jpsi\to\ppbar\phi$ is studied via the two decay modes
$\phi\to\kskl$ and $\phi\to\kk$ using a data sample of
$1.31\times10^{9}$ $\jpsi$ events~\cite{njpsi09,njpsi12} accumulated
with the \mbox{BESIII} detector.

\section{\boldmath BESIII Detector and Monte Carlo Simulation}

The BESIII detector~\cite{detector} is a general purpose spectrometer at
the BEPCII $e^+e^-$ accelerator for studies of hadron spectroscopy as well as
$\tau$-charm physics~\cite{tcphy}. The BESIII detector with a
geometrical acceptance of 93\% of 4$\pi$ consists of the following
main components: 1) a small-cell, helium-based main drift chamber
(MDC) with 43 layers, which measures tracks of charged particles and
provides a measurement of the specific energy loss $dE/dx$. The
average single wire resolution is 135~$\mu$m, and the momentum
resolution for 1~GeV/$c$ charged particles in a 1~T magnetic field is
0.5\%;
2) a Time-Of-Flight system (TOF) for particle
identification (PID) composed of a barrel part constructed of two
layers with 88 pieces of 5~cm thick, 2.4~m long plastic scintillators
in each layer, and two end caps with 48 fan-shaped, 5~cm thick,
plastic scintillators in each end cap.  The time resolution is 80~ps
(110~ps) in the barrel (end caps), corresponding to a $K/\pi$
separation of more than $2\sigma$ for momenta at 1~GeV/$c$ and below;
3) an electromagnetic calorimeter (EMC) consisting of 6240 CsI(Tl) crystals
arranged in a cylindrical shape (barrel) plus two end caps.
For 1~GeV/$c$ photons, the energy resolution is 2.5\% (5\%) in
the barrel (end caps), and the position resolution is 6~mm (9~mm) in
the barrel (end caps);
4) a muon chamber system (MUC) consists of about 1200~m$^{2}$ of Resistive
Plate Chambers (RPC) arranged in 9 layers in the barrel and 8 layers
in the end caps and incorporated in the return iron yoke of the
superconducting magnet. The position resolution is about 2~cm.

The optimization of the event selection, the determination of the
detector efficiency and the estimation of backgrounds are performed
through Monte Carlo (MC) simulations.
The \textsc{geant}{\footnotesize 4}-based~\cite{geant4} simulation software
\textsc{boost}~\cite{liu5} includes the geometric and material
description of the BESIII detectors and models for the detector response and
digitization, as well as the tracking of the detector running
conditions and performance. For the background study, an inclusive MC
sample of $1.23\times10^{9}$ $\jpsi$ decay events is generated. The
production of the $\jpsi$ resonance is simulated by the MC event
generator \textsc{kkmc}~\cite{liu61,liu62}, while the decays are
generated by \mbox{\textsc{evtgen}}~\cite{liu7} for known decay modes
with branching fractions being set to Particle Data Group (PDG) world
average values~\cite{PDG}, and by \textsc{lundcharm}~\cite{liu9} for
the remaining unknown decays. A sample of $2.0\times10^{5}$ events is generated
for the three-body decay $\jpsi\to\ppbar\phi$ using a flat distribution in phase
space (PHSP), and the signal detection efficiency is obtained by weighting the
PHSP MC to data. For the decay $\jpsi\to X(\ppbar)\phi\to\ppbar\phi$, a sample
of $2.0\times10^{5}$ events is generated, and the angular distribution is
considered in the simulation.

\section{\boldmath Event selection and background analysis}

Two dominant $\phi$ decays are used to reconstruct the $\phi$ meson in
the study of the decay $\jpsi\to\ppbar\phi$, which allows us to check
our measurements and to improve the precision of our results. In the
following text, if not special specified, $K\bar{K}$ refers to both
$\kskl$ and $\kk$ final states.

\subsection{\boldmath $J/\psi\to\ppbar\phi$, $\phi\to\kskl$}

In this decay channel, the $\ks$ is reconstructed in its decay to two
charged pions, while the long-lived, difficult to detect, $\kl$ is
taken as a missing particle. The event topology is therefore
$\ppbar\pp\kl$, and candidate events must have at least four charged
tracks.  Each of the charged track is reconstructed from MDC hits and
the polar angle $\theta$ must satisfy $|\cos\theta|<0.93$.

Two of the charged tracks are identified as proton and anti-proton by
using combined TOF and $dE/dx$ information, while all other tracks are
assumed to be charged pions without PID requirement. The identified proton and
anti-proton are further required to originate from the same primary vertex and
pass within 10~cm in the beam direction and within 1~cm in the radial
direction with respect to the interaction point.

\begin{figure}[htbp]
\centering
\begin{overpic}[width=8.0cm,height=5.0cm,angle=0]{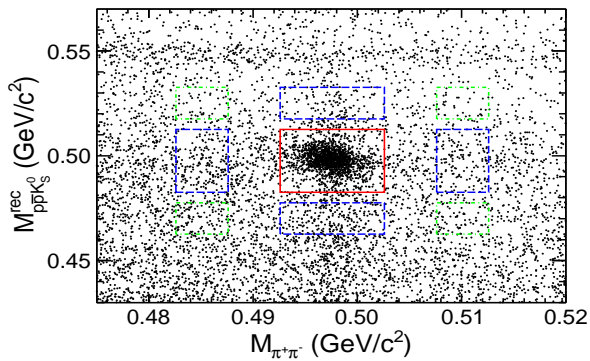}
\end{overpic}
\vskip -0.0cm
\parbox[1cm]{8cm}{
  \caption{Scatter plot of the $\pp$ invariant
    mass versus the recoiling mass against $\ppbar\ks$; the boxes
    represent the $\ks$ and $\kl$ signal region and sideband regions described
    in the text.}
 \label{scatter}}
\end{figure}

The $\ks$ meson is reconstructed by constraining a pair of oppositely
charged pions to originate from a secondary vertex, and only candidate
events with only one successfully reconstructed $\ks$ candidate are preserved
for the further analysis. To suppress backgrounds, the chi-square of
the second vertex fit is required to be less than 40. The scatter plot
of the $\pp$ invariant mass ($M_{\pp}$) versus the recoiling mass
against $\ppbar\ks$ ($M^\text{rec}_{\ppbar\ks}$) is shown in
Fig.~\ref{scatter}, where a prominent $\ks-\kl$ cluster corresponding
to the signal channel of $\jpsi\to\ppbar\kskl$ is observed. Mass
windows of $|M_{\pp}-m_{K^0}|<$5~MeV/c$^{2}$ and
$|M^\text{rec}_{\ppbar\ks}-m_{K^0}|<$15~MeV/c$^{2}$ are required to
identify signal events, where $m_{K^0}$ is the nominal mass of $K^0$
from PDG~\cite{PDG}.

\begin{figure*}[htbp]
\centering
\begin{overpic}[width=7.0cm,height=5.0cm,angle=0]{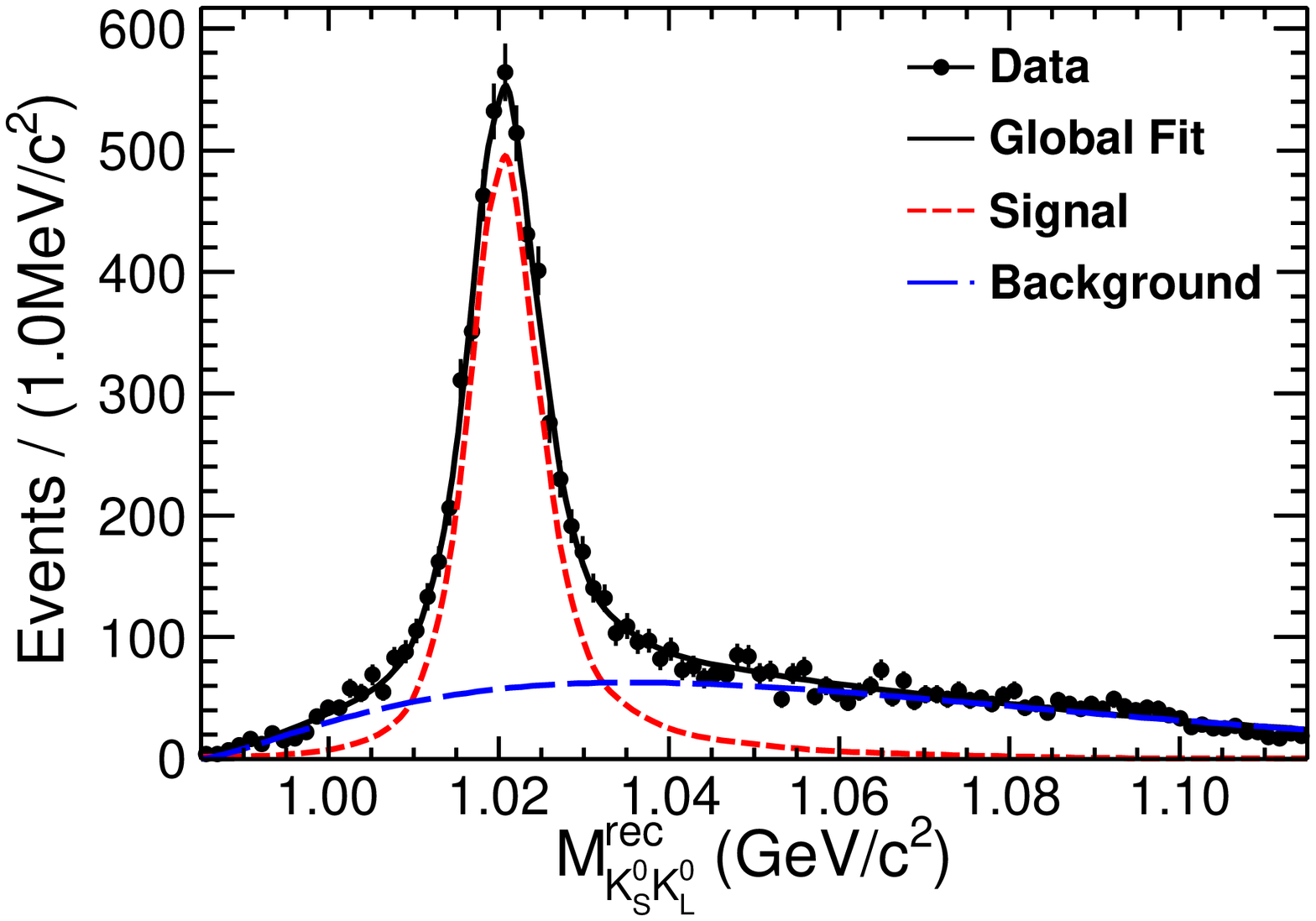}
\put(20,58){\large\bf (a)}
\end{overpic}
\begin{overpic}[width=7.0cm,height=5.0cm,angle=0]{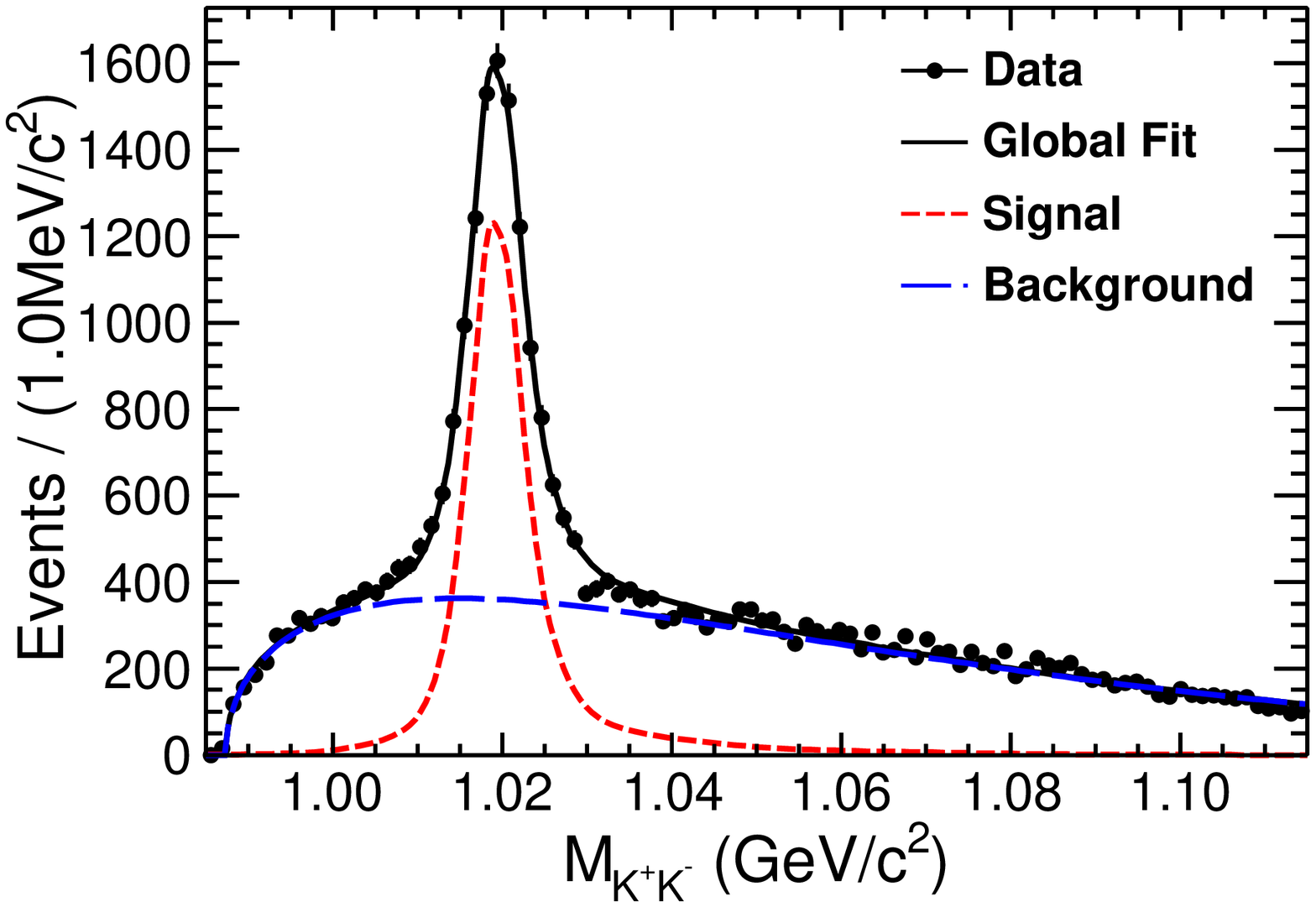}
\put(20,58){\large\bf (b)}
\end{overpic}
\vskip -0.0cm
\parbox[1cm]{14cm}{
  \caption{ Fits to (a) the recoil mass spectrum against the
    $\ppbar$ system of the $\ppbar\kskl$ candidates and (b) the
    $K^+K^-$ invariant mass spectrum of the $\ppbar\kk$
    candidates. The black solid lines are the global fit results, the
    short dashed lines are the signal shapes, and the long dashed
    lines represent the background shapes.}
\label{phim}}
\end{figure*}

After applying the previously mentioned selection criteria, the recoil
mass against the $\ppbar$ system, $M^\text{rec}_{\ppbar}$, is examined, as
shown in Fig.~\ref{phim} (a), in which a clear $\phi$ signal is
observed. To estimate the combinational backgrounds from non-$\ks$ or
non-$\kl$ events, the background events in the $\ks$ and $\kl$
sideband regions, as indicated in Fig.~\ref{scatter}, are
investigated. More specifically, the sideband ranges are defined as
10~MeV/$c^2$$<|M_{\pp}-m_{\ks}|<$
15~MeV/$c^2$
and 20~MeV/$c^2<|M^\text{rec}_{\ppbar\ks}-m_{\kl}|<$
35~MeV/$c^2$.
The sideband events do not form a peaking background around the
$\phi$ nominal mass in the $M^\text{rec}_{\ppbar}$
spectrum. In addition, the other background sources are examined by
analyzing the inclusive MC sample of $\jpsi$
decay. The potential background contributions from the inclusive MC
sample are found to be the channels with $\ppbar\pp\piz\piz$
final states, such as $\jpsi\to\ppbar
f^{\prime}_{0}\to\ppbar\ks\ks$, and $\jpsi\to
p\omega\bar{\Delta}^{-}+c.c.$, but none of these backgrounds produce a
peak around the $\phi$ nominal mass.

\subsection{\boldmath $J/\psi\to\ppbar\phi$, $\phi\to\kk$}

For $\jpsi\to\ppbar\phi$ with $\phi\to\kk$, the final states are
$\ppbar\kk$. Since the $\ppbar\phi$ mass threshold is close to the
$\jpsi$ nominal mass, the available kinematic energy for the kaons is
small in this reaction. As a consequence, one of the two charged kaons
will have a relatively low momentum and is, thereby, difficult to
reconstruct. Therefore, the candidate events are required to have
three or four charged tracks. The selection criteria for the charged
tracks are same as for the proton (anti-proton) as described in the
previous subsection. Two of the charged tracks are required to be
identified as proton and anti-proton, while the others are required to be
identified as kaons.

A one-constraint (1C) kinematic fit is applied in which the
missing mass of the undetected kaon is constrained to its nominal
mass. In the case where both kaons have been detected, two 1C kinematic fits
are performed with the missing $K^+$ or $K^-$ assumptions, and the one
with the smallest chi-square is retained. To suppress backgrounds, the
chi-square of 1C kinematic fit is required to be less than 10.

After the above selection criteria, the background contamination is
investigated using the inclusive $\jpsi$ MC sample. Besides the
irreducible backgrounds from non-resonant $\jpsi\to\ppbar\kk$, the reducible background
is evaluated to be 20\% of all selected events, dominated by
the processes involving $\Lambda$ ($\bar{\Lambda}$) intermediate
states. To suppress the above backgrounds, all other charged tracks except for
the selected proton, antiproton and kaon candidates are
assumed to be pions, and the events are vetoed if any combination of
$p\pi^-$ or $\bar{p}\pi^+$ has an invariant mass lying in the range
$|M_{p\pim(\bar{p}\pip)}-M_{\Lambda(\bar{\Lambda})}|<$
10~MeV/$c^{2}$. The $\Lambda$ ($\bar{\Lambda}$) veto requirement
retains about 97\% of the signal events while rejecting about
two-thirds of corresponding reducible backgrounds.

The $K^+K^-$ invariant mass distribution after applying all the above
mentioned selection criteria is shown in Fig.~\ref{phim} (b). A clear
$\phi$ peak, corresponding to the signal of $\jpsi\to\ppbar\phi$,
is observed.  Using the inclusive $\jpsi$ MC sample, the main
backgrounds are found to be the processes of
$\jpsi\to\Lambda(1520)\bar{\Lambda}(1520)$ and
$\jpsi\to pK^{-}\Lambda(1520)+c.c.$ with $\Lambda(1520)\to pK$. These
processes can be seen in the data as well, but none of these
backgrounds contribute to the $\phi$ peak.

\section{\boldmath Measurement of $\mathcal{B}(\jpsi\to\ppbar\phi)$}

\begin{figure*}[htbp]
\centering
\begin{overpic}[width=17.0cm,height=7.0cm,angle=0]{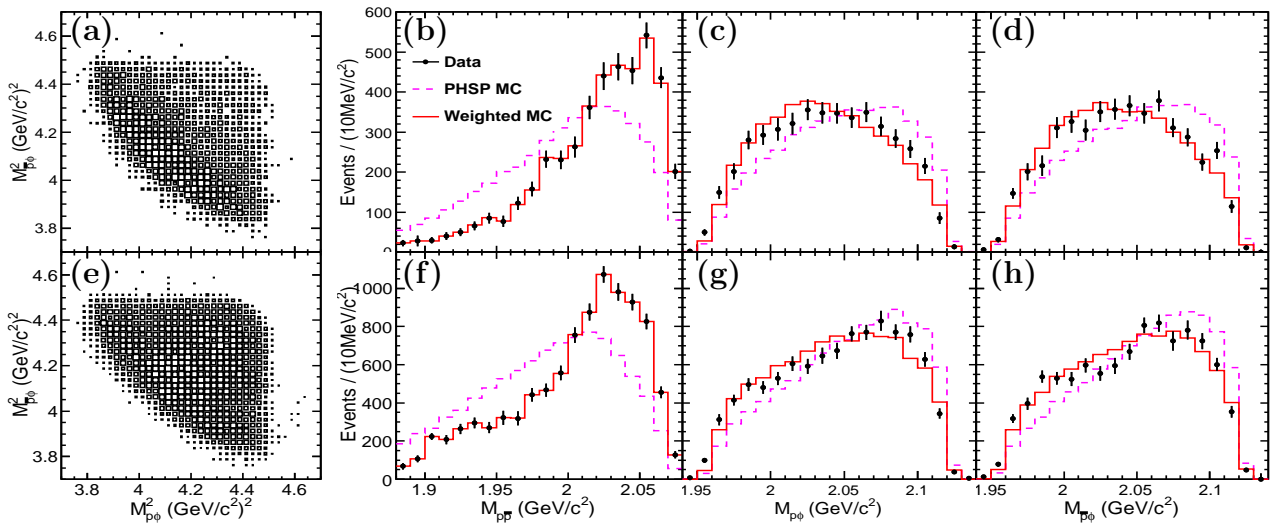}
\put(5,38){\large\bf (a)}
\put(31,38){\large\bf (b)}
\put(54,38){\large\bf (c)}
\put(77,38){\large\bf (d)}
\put(5,19){\large\bf (e)}
\put(31,19){\large\bf (f)}
\put(54,19){\large\bf (g)}
\put(77,19){\large\bf (h)}
\end{overpic}
\vskip -0.0cm
\hskip  0.0cm
\parbox[1cm]{17cm}{
\caption{Dalitz plots of the data and the $\ppbar$, $p\phi$, and $\bar{p}\phi$ invariant masses. The upper row (a, b, c, d) and the lower row (e, f, g, h) correspond to $\phi\to\kskl$ and $\phi\to\kk$, respectively. The dots with error bars represent the background-subtracted data, the dashed histograms represent the PHSP MC simulations, and the solid histograms represent the reweighted MC simulation.}
\label{dist}}
\end{figure*}

The signal yields of $\jpsi\to\ppbar\phi$ for the two decay modes are
obtained from unbinned maximum likelihood fits to the
$M^\text{rec}_{\ppbar}$ and $M_{K^+K^-}$ mass spectra. In the fit of each
mode, the $\phi$ signal is described by the line shape obtained from
the MC simulation convoluted with a Gaussian function, which accounts
for the difference of mass resolution between the data and the MC. The
background shape is parameterized by an ARGUS function~\cite{argus}.
The parameters of the Gaussian function and the ARGUS function are left free in the fit.
The projections of the fits are shown in Fig.~\ref{phim}, and the signal yields are listed in
Table.~\ref{branchfraction}.

The detection efficiencies are obtained by MC simulations that
are, in the first instance, based on a PHSP three-body decay of the signal
mode $\jpsi\to\ppbar\phi$.  However, it is found that data deviate
strongly from the PHSP MC distributions, as the histograms shown in Fig.~\ref{dist},
where, to subtract the backgrounds, the signal yields of data in each
bin are extracted by fitting the $\phi$ signal in the $\kkb$ invariant
mass. The detection efficiency varies significantly at low
momenta of proton and anti-proton, and, therefore, strongly depends
on the $\ppbar$ invariant mass. To obtain a more accurate detection
efficiency, the events of the PHSP MC are weighted according to the
observed $\ppbar$ mass distribution, where the weight factor is the
ratio of $\ppbar$ mass distributions between data and the PHSP MC in
Fig.~\ref{dist} (b) and (f). The average detection efficiencies are
determined to be $(30.8\pm0.2)$\% and $(28.9\pm0.1)$\% for
$\phi\to\ks\kl$ and $\phi\to K^+K^-$, respectively. The weighted PHSP
MC distributions of the $\ppbar$, $p\phi$ and $\bar{p}\phi$ invariant
masses are approximately consistent with the background-subtracted
data, as shown by the solid lines in Fig.~\ref{dist}. As for
the small discrepancies between the weighted PHSP MC and the data,
a secondary reweighting is performed based on the present results,
and the difference is considered as a systematic uncertainty.

The branching fraction of $\jpsi\to\ppbar\phi$ is
calculated using
\begin{equation}
 \label{Brppphi}
 \mathcal{B}(\jpsi\to\ppbar\phi)
 =\frac{N_\text{obs}}{N_{\jpsi}\times\varepsilon\times\mathcal{B}(\phi\to K\bar{K})},
\end{equation}
where $N_\text{obs}$ is the number of signal yields from the fit,
$N_{\jpsi} = (1.31\pm0.01)\times 10^{9}$ is the total number of $\jpsi$
events~\cite{njpsi12} determined from $\jpsi$
inclusive decays, $\varepsilon$ is the weighted detection efficiency
obtained as described above, and $B(\phi\to K\bar{K})$ represents the
branching fraction of $\phi\to\ks\kl$ or $\phi\to K^+K^-$, taking into
account the branching fraction of $\ks\to\pp$.

The branching fractions of $\jpsi\to\ppbar\phi$ measured using the two
$\phi$ decay modes are summarized in Table~\ref{branchfraction}. The
results are consistent with each other within statistical
uncertainties. These two branching fractions are combined using a
weighted least-square approach~\cite{combine}, where the systematic
uncertainties on the tracking and PID efficiencies of proton and
anti-proton as well as the number of $\jpsi$ events are common for
the two decay modes, and the remaining systematic uncertainties are independent for each mode.
The systematic uncertainties are discussed in detail in the next
section. The combined branching fraction, $\mathcal{B}(\jpsi\to\ppbar\phi)$, is calculated
to be $(5.23\pm0.06\pm0.33) \times 10^{-5}$, where the first
uncertainty is the statistical and the second systematic.

\begin{table}[htbp]
\begin{center}
{\caption {Signal yields, weighted detection efficiencies and the branching
    fractions of $\jpsi\to\ppbar\phi$ measured by the two decay modes. The first
    errors are statistical and the second systematic (see
    Section~\ref{syserr}).}
  \label{branchfraction}}
\begin{tabular}{lccc}
\hline \hline
$\phi$ decay mode  &  $N_\text{obs}$  &  $\varepsilon$(\%)  &        $\mathcal{B}(\jpsi\to\ppbar\phi)$         \\ \hline
$\phi\to\kskl$     &   4932$\pm$101   &      30.8$\pm$0.2   &   (5.17$\pm$0.11$\pm$0.44)$\times$10$^{-5}$      \\
$\phi\to\kk$       &   9729$\pm$148   &      28.9$\pm$0.1   &   (5.25$\pm$0.08$\pm$0.43)$\times$10$^{-5}$      \\
\hline \hline
\end{tabular}
\end{center}
\end{table}

\section{\boldmath Systematic uncertainties}
\label{syserr}
The systematic uncertainties are estimated by taking into account the
differences in efficiencies between data and MC for the tracking and
PID algorithms, the $\ks$ reconstruction, the $\ks/\kl$ mass window
requirement, the kinematic fit and the $\Lambda$ ($\bar{\Lambda}$)
veto. In addition, the uncertainties associated with the mass spectrum
fit, the weighting procedure, as well as the branching fraction of the
intermediate state decay and the total number of $\jpsi$ events are
taken into consideration.

\begin{itemize}
\item[1)] {\bf MDC tracking:} the MDC tracking efficiencies of
  $p/\bar{p}$ and $K^\pm$ are measured using clean samples of
  $\jpsi\to\ppbar\pp$ and
  $\jpsi\to\ks K^{\pm}\pi^{\mp}$~\cite{tracking,tracking2},
  respectively. The difference in tracking efficiencies between data
  and MC is 1.2\% for protons, 1.9\% for antiprotons, and 1.0\% for
  kaons. The systematic uncertainty associated with the tracking
  efficiency of $\pi^\pm$ is included in the uncertainty of $\ks$
  reconstruction.

\item[2)] {\bf PID efficiency:} To estimate the PID efficiency
  uncertainty, we study $p/\bar{p}$ and $K^\pm$ PID efficiencies with
  the same control samples as those used in the tracking
  efficiency. The average PID efficiency difference between data and
  MC is found to be 2\% per charged track and taken as a systematic
  uncertainty.

\item[3)] {\bf $\ks$ reconstruction:} the $\ks$ reconstruction
  involves the charged-track reconstruction of the $\pip\pim$ pair and
  a second vertex fit. The corresponding systematic uncertainty is
  estimated using a control sample of the decay
  $\jpsi\to\phi K^{0}_{S}K^{\pm}\pi^{\mp}$. The relative difference in
  the reconstruction efficiencies of the $\ks$ between data and MC is
  4.2\% and taken as a systematic uncertainty.

\item[4)] {\bf $\ks$ and $\kl$ mass window:} Due to the difference in
  the mass resolutions between data and MC, the uncertainty related
  with the $\ks$ or $\kl$ mass window requirement is investigated by
  smearing the MC simulation in accordance with the signal shape of
  data. The changes on the detection efficiencies, $1.3\%$ and
  $2.5\%$, are assigned as the systematic uncertainties for the $\ks$
  and $\kl$ mass window requirements, respectively.

\item[5)] {\bf 1C kinematic fit:} To estimate the systematic
  uncertainty from the 1C kinematic fit, a clean control sample
  $\jpsi\to pK^{-}\bar{\Lambda} +c.c.$ is selected without using a
  kinematic fit. The efficiency of 1C kinematic fit is estimated by
  the ratio of signal yields with ($\chi^{2}_{1C}<$~10 required) and
  without 1C kinematic fit. The corresponding difference in the
  efficiencies between data and MC is found to be 1.4\% and taken as a
  systematic uncertainty.

\item[6)] {\bf $\Lambda/\bar{\Lambda}$ veto:} the requirement
  $|M_{p\pim/\bar{p}\pip}-M_{\Lambda/\bar{\Lambda}}|>$ 10~MeV$/c^{2}$
  is applied to veto $\Lambda/\bar{\Lambda}$ background events. The
  alternative choices
  $|M_{p\pim/\bar{p}\pip}- M_{\Lambda/\bar{\Lambda}}|$
  $>5$~MeV$/c^{2}$, or $>$ 15~MeV$/c^{2}$ are implemented to
  recalculate the branching fraction. The maximum difference of the
  final results, 0.6\%, is taken as a systematic uncertainty.

\item[7)] {\bf Mass spectrum fit:} The systematic uncertainty
  associated with the fit of the mass spectrum comes from the
  parameterization of the signal shape, the background shape and the
  fit range. To estimate the uncertainty from the $\phi$ signal shape,
  we perform an alternative fit with an acceptance corrected
  Breit-Wigner to describe the $\phi$ signal shape. The uncertainty
  associated with the smooth shape of the background underneath the
  $\phi$ peak is evaluated by replacing the ARGUS function with a
  function of $f(M)=(M-M_a)^{c}(M_b-M)^{d}$, where, $M_a$ and $M_b$
  are the lower and upper edges of the mass distribution,
  respectively; $c$ and $d$ are free parameters. The uncertainty due
  to the fit range is estimated by fitting within the alternative
  ranges. The change of signal yield in the different fit scenarios is
  taken as the corresponding systematic uncertainty. The quadratic
  sums of the three individual uncertainties, 3.9\% and 1.9\%, for
  $\phi\to\kskl$ and $\phi\to\kk$, respectively, are taken as the
  systematic uncertainty related with the mass spectrum fit.

\item[8)] {\bf Weighting procedure:} To obtain a reliable detection
  efficiency, the PHSP MC sample is weighted to match the distribution
  of the background-subtracted data. To consider the effect on the
  statistical fluctuations of the signal yield in the data, a set of
  toy-MC samples, which are produced by sampling the signal yield and
  its statistical uncertainty of the data in each bin, are used to
  estimate the detection efficiencies. Consider the systematic
  uncertainty on the secondary reweighting, the resulting deviations of
  detection efficiencies, 2.4\% and 2.9\% for $\phi\to\kskl$ and
  $\phi\to\kk$, respectively, are taken as the systematic uncertainty
  associated with the weighting procedure.

\end{itemize}

The contributions of the systematic uncertainties from the above
sources and the systematic uncertainties of the branching fractions of
intermediate decays ($\phi\to\kk$ and $\ks\to\pi^+\pi^-$) as well as
the number of $J/\psi$ events~\cite{njpsi09,njpsi12} are summarized in
Table~\ref{systemtotal}. The total systematic uncertainties are given
by the quadratic sum of the individual uncertainties, assuming all
sources to be independent.

\begin{table*}[hbtp]
  \centering
  {\caption {Summary of the systematic uncertainties in the branching fraction
      measurement (in \%), the items with ... denote that the corresponding systematic
      uncertainty is not applicable.}
\label{systemtotal}}
\begin{tabular}{lcccc}
\hline
\hline
Sources         &              \multicolumn{2}{c}{$\phi\to\kskl$}   &   \multicolumn{2}{c}{$\phi\to\kk$}     \\
\cline{2-5}
                &$\mathcal{B}(\jpsi\to\ppbar\phi)$& $\mathcal{B}(\jpsi\to X(\ppbar)\phi\to\ppbar\phi)$
                &$\mathcal{B}(\jpsi\to\ppbar\phi)$& $\mathcal{B}(\jpsi\to X(\ppbar)\phi\to\ppbar\phi)$       \\ \hline
MDC tracking              &      3.1        &           3.1         &          4.1      &          4.1       \\
PID efficiency            &      4.0        &           4.0         &          6.0      &          6.0       \\
$\ks$ reconstruction      &      4.2        &           4.2         &          ...      &          ...       \\
$\ks$ mass window         &      1.3        &           1.3         &          ...      &          ...       \\
$\kl$ mass window         &      2.5        &           2.5         &          ...      &          ...       \\
1C kinematic fit          &      ...        &           ...         &          1.4      &          1.4       \\
$\Lam$($\Lamb$) veto      &      ...        &           ...         &          0.6      &          0.6       \\
Mass spectrum fit         &      3.9        &           ...         &          1.9      &          ...       \\
Weighting procedure       &      2.4        &           ...         &          2.9      &          ...       \\
Number of $\jpsi$ events  &      0.8        &           0.8         &          0.8      &          0.8       \\
$\mathcal{B}(\phi\to\kkb)$&      1.2        &           1.2         &          1.0      &          1.0       \\
$\mathcal{B}(\ks\to\pp)$  &      0.1        &           0.1         &          ...      &          ...       \\
Total                     &      8.6        &           7.3         &          8.3      &          7.5       \\ \hline \hline
\end{tabular}
\end{table*}

\section{\boldmath Upper limit of $\ppbar$ mass threshold enhancement }

The Dalitz plots of the data and the corresponding one-dimensional
mass projections presented in Fig.~\ref{dist} show no significant signatures
of a threshold enhancement in the $\ppbar$ invariant mass
nor obvious structures in the $p\phi$ ($\bar{p}\phi$) mass spectra.
The most rigorous procedure is to carry out a PWA. However, due to
the small phase space for the decay $\jpsi\to\ppbar\phi$ and the lack
of a proper physics model, such an analysis is difficult to pursue.
In this analysis, we only consider an upper limit for the $\ppbar$ mass
threshold enhancement by fitting solely the $\ppbar$ mass spectrum
near the threshold.

To obtain the best upper limit on the $X(\ppbar)$ yield, the two decay
modes are combined to determine the upper limit on the branching
fraction of $\jpsi\to X(\ppbar)\phi\to\ppbar\phi$. A least squares
simultaneous fit is performed on both $\ppbar$
invariant mass distributions of the two $\phi$ decay modes around the mass
threshold. The two decay modes share the same branching fraction
\begin{equation}
\mathcal{B}=\frac{N_\text{obs}}{N_{\jpsi}\cdot\mathcal{B}(\phi\to K\bar{K})\cdot\varepsilon\cdot(1-\sigma_\text{sys})},
\end{equation}
where $N_\text{obs}$ represents the $X(\ppbar)$ signal yield of each decay
mode corresponding to the given test
$\mathcal{B}(\jpsi\to X(\ppbar)\phi\to\ppbar\phi)$, $N_{\jpsi}$ and
$\mathcal{B}(\phi\to K\bar{K})$ are same as described in
Eq.~\ref{Brppphi}, $\varepsilon$ is the detection efficiency of
$X(\ppbar)$ obtained from MC simulations (14.4\% for the mode
$\phi\to\kskl$, while 21.4\% for $\phi\to\kk$), $\sigma_\text{sys}$ is the
total relative systematic uncertainty as reported in
Table~\ref{systemtotal}. With such a method, a combined upper limit on
the branching fraction, $\mathcal{B}^{\text{UL}}$, at a 90\% C.L. can be
determined directly.

In the simultaneous fit, the spin and parity of $X(\ppbar)$ are set to be
$0^{-+}$ based on earlier BESIII observations~\cite{BESIII2}, and
effects of interference are neglected. The signal of $X(\ppbar)$ is
parameterized by an acceptance-weighted $\mathcal{S}$-wave
Breit-Wigner function
\begin {equation}
  BW(M)\simeq \frac{f_\text{FSI}\times q^{2L+1}\kappa^{3}}{(M^{2}-M^{2}_{0})^{2}+M^{2}_{0}\Gamma^{2}_{0}}
           \times\varepsilon_\text{rec}(M),
  \label{SBW}
\end {equation}
where $M$ is the $\ppbar$ invariant mass, $q$ is the momentum of the
proton in the $\ppbar$ rest frame, $\kappa$ is the momentum of the
$\phi$ in the $\jpsi$ rest frame, $L=0$ is the relative orbital
angular-momentum of $\ppbar$ system, $M_0$ and $\Gamma_0$ are the mass
and width of the $X(\ppbar)$~\cite{BESIII2}, $\varepsilon_\text{rec}(M)$ is
the detection efficiency as a function of $\ppbar$ invariant mass,
which is obtained from the MC simulations of
$\jpsi\to X(\ppbar)\phi\to\ppbar\phi$ by taking into account the
helicity angular distributions, the parameter $f_\text{FSI}$ accounts for
the effect of the FSI.

\begin{figure}[htbp]
\centering
\begin{overpic}[width=8.5cm,height=4.0cm,angle=0]{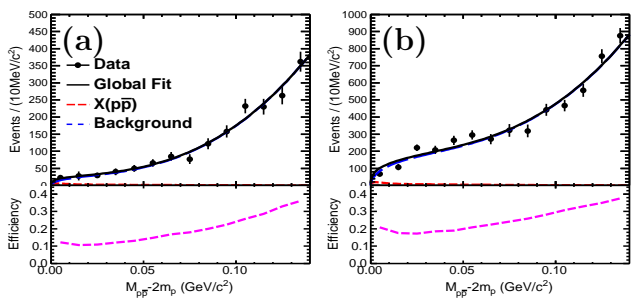}
\put(10,40){\large\bf (a)}
\put(60,40){\large\bf (b)}
\end{overpic}
\vskip -0.0cm
\parbox[1cm]{8.5cm} {
\caption{Distributions of $M_{\ppbar}-2m_{p}$ and the fit results corresponding to the upper limit on the branching fraction at the 90\% C.L., the dashed line at the bottom is the efficiency as a function of the $\ppbar$ mass, (a) for $\phi\to\kskl$, (b) for $\phi\to\kk$. }
\label{uplimitfit}}
\end{figure}

To take into account the systematic uncertainties related to the fit
procedure of the $X(\ppbar)$, three aspects with different fit
scenarios are considered: (1) excluding the FSI factor (corresponding
to $f_\text{FSI}$=1); taking into account the J\"ulich FSI value for
FSI~\cite{ppfsi2}; (2) the non-resonant backgrounds both parameterized
by a function of
$f(\delta)=N(\delta^{1/2}+a_{1}\delta^{3/2}+a_{2}\delta^{5/2})$
($\delta=M_{\ppbar}-2m_p$, $m_p$ is the proton mass, $a_{1}$ and
$a_{2}$ are free parameters); or both represented by the shape
obtained from the $\jpsi\to\ppbar\phi$ MC simulation; (3) the fit
ranges both in [0.0, 0.140] or in [0.0, 0.150] GeV/c$^{2}$. By
combining these three different aspects, we perform in total eight
alternative fit scenarios. The fit scenario taking into account
the FSI, with the non-resonant backgrounds parameterized by the
function, and the fit ranges both in [0.0, 0.140] GeV/c$^{2}$, gives
the maximum upper limit on the branching fraction, which is shown in
Fig.~\ref{uplimitfit}, where the efficiency as a function of the
$\ppbar$ mass is also plotted. The combined upper limit at the 90\%
C.L.\ is determined to be $2.1\times10^{-7}$.

\section{\boldmath Summary}

Based on a sample of 1.31$\times$10$^{9}$ $\jpsi$ events accumulated
at BESIII, we present a study of $\jpsi\to\ppbar\phi$ with two decay
modes $\phi\to\ks\kl$ and $\phi\to\kk$. The branching fraction of
$\jpsi\to\ppbar\phi$ is measured to be
$[5.23\pm0.06\stat\pm0.33\syst]\times10^{-5}$, which is consistent
with the previous measurement~\cite{DM2}, but with a significantly
improved precision. We have neither observed a significant structure
in the $p\phi$ or $\bar{p}\phi$ mass spectra, nor found evidence of an
enhancement in the $\ppbar$ mass spectrum near its threshold. The
corresponding upper limit on the branching fraction of
$\jpsi\to X(\ppbar)\phi\to\ppbar\phi$ is determined to be
$2.1\times 10^{-7}$ at a 90\% C.L.. With the production branching
fraction of $\jpsi\to\gamma X(\ppbar)\to\gamma\ppbar$,
$[9.0^{+0.4}_{-1.1}\stat^{+1.5}_{-5.0}\syst\pm2.3\model]\times10^{-5}$~\cite{BESIII2},
the upper limit on the decay rate ratio of
$\mathcal{B}(\jpsi\to X(\ppbar)\phi)/\mathcal{B}(\jpsi\to \gamma
X(\ppbar))$
is calculated to be
$[0.23^{+0.01}_{-0.03}\stat^{+0.04}_{-0.13}\syst\pm0.06\model]\%$.

Though no clear structure in the $\ppbar$, $p\phi$ and $\bar{p}\phi$
mass spectra is observed in this analysis, the data appear to
significantly deviate from a naive PHSP distribution. This implies the
existence of interesting dynamical effects, such as intermediate
resonances. With the presented analysis, it is difficult to study them
in detail due to the small phase space of the decay
$\jpsi\to\ppbar\phi$. The study of analogous decay processes with
larger phase space, such as $\psi(3686)\to\ppbar\phi$, in combination
with a PWA, may shed light and help to understand their dynamical
origins.

\section{\boldmath ACKNOWLEDGMENTS}

The BESIII collaboration thanks the staff of BEPCII and the IHEP
computing center for their strong support. This work is supported in
part by National Key Basic Research Program of China under Contract No.
2015CB856700; National Natural Science Foundation of China (NSFC) under
Contracts Nos. 11125525, 11235011, 11322544, 11335008, 11425524,
11375170, 11275189, 11475169, 11475164, 11175189;
the Chinese Academy of Sciences (CAS) Large-Scale Scientific Facility
Program; the CAS Center for Excellence in Particle Physics (CCEPP); the
Collaborative Innovation Center for Particles and Interactions (CICPI);
Joint Large-Scale Scientific Facility Funds of the NSFC and CAS under
Contracts Nos. 11179007, U1532102, U1232201, U1332201; CAS under Contracts Nos.
KJCX2-YW-N29, KJCX2-YW-N45; 100 Talents Program of CAS; National 1000
Talents Program of China; INPAC and Shanghai Key Laboratory for Particle
Physics and Cosmology; German Research Foundation DFG under Contract No.
Collaborative Research Center CRC-1044; Istituto Nazionale di Fisica
Nucleare, Italy; Koninklijke Nederlandse Akademie van Wetenschappen
(KNAW) under Contract No. 530-4CDP03; Ministry of Development of Turkey
under Contract No. DPT2006K-120470; Russian Foundation for Basic
Research under Contract No. 14-07-91152; The Swedish Resarch Council; U.
S. Department of Energy under Contracts Nos. DE-FG02-05ER41374,
DE-SC-0010504, DE-SC0012069, DESC0010118; U.S. National Science
Foundation; University of Groningen (RuG) and the Helmholtzzentrum fuer
Schwerionenforschung GmbH (GSI), Darmstadt; WCU Program of National
Research Foundation of Korea under Contract No. R32-2008-000-10155-0.

\end{document}